\documentclass[aps,twocolumn,preprintnumbers,floatfix,nofootinbib]{revtex4}

\usepackage{graphicx}

\usepackage{amsmath}
\usepackage{amsfonts}
\usepackage{amssymb}
\usepackage{cleveref}
\usepackage[utf8]{inputenc}
\usepackage{color}

\def\be{\begin{equation}}
\def\ee{\end{equation}}
\def\bea{\begin{eqnarray}}
\def\eea{\end{eqnarray}}

\def\vep{\varepsilon}

\def\<{\langle}
\def\>{\rangle}

\def\mt{{\mathcal{T}}}

\def\vep{\varepsilon}

\def\mt{{\mathcal{T}}}

\def\slashchar#1{\setbox0=\hbox{$#1$}           
   \dimen0=\wd0                                 
   \setbox1=\hbox{/} \dimen1=\wd1               
   \ifdim\dimen0>\dimen1                        
      \rlap{\hbox to \dimen0{\hfil/\hfil}}      
      #1                                        
   \else                                        
      \rlap{\hbox to \dimen1{\hfil$#1$\hfil}}   
      /                                         
   \fi}          

\def\mI{\mathcal{I}}

\def\mL{\mathcal{L}}

\def\mO{\mathcal{O}}

\def\bea{\begin{eqnarray}}
\def\eea{\end{eqnarray}}
\def\beq{\begin{equation}}
\def\eeq{\end{equation}}
\def\vep{\varepsilon}

\def\<{\langle}
\def\>{\rangle}

\def\ewd{{\mathcal{D}}}

\newcommand{\Eq}[1]{Eq.~(\ref{#1})}

\begin{document}

\preprint{CERN-TH-2018-100}

\title{Universal Imprints of a Pseudo-Nambu-Goldstone Higgs Boson}

\author{Da Liu$^{\, a}$, Ian Low$^{\, a,b,c}$ and Zhewei Yin$^{\, b}$}
\affiliation{
\mbox{$^a$ High Energy Physics Division, Argonne National Laboratory, Argonne, IL 60439, USA}\\
\mbox{$^b$ Department of Physics and Astronomy, Northwestern University, Evanston, IL 60208, USA} \\
\mbox{$^c$ Theoretical Physics Department, CERN, 1211 Geneva 23, Switzerland}
}

\begin{abstract}
A large class of models addressing the electroweak naturalness problem postulates  the existence of new spontaneously broken global symmetries above the weak scale. The Higgs boson  arises as a pseudo-Nambu-Goldstone boson (pNGB) whose interactions are nonlinear due to the presence of degenerate vacua.  We argue that, once the normalization of the pNGB decay constant $f$ is determined, the Higgs nonlinear interactions  in the gauge sector are universal in the infrared and independent of the symmetry breaking pattern $G/H$,  even after integrating out heavy composite resonances.  We  propose a set of "universal relations" in Higgs couplings with electroweak gauge bosons and in triple gauge boson couplings, which are unique predictions of the universal nonlinearity. Experimental measurements of these relations would serve as the litmus test of a pNGB Higgs boson.
\end{abstract}


\maketitle

\noindent
{\bf Introduction.}--The discovery of the 125 GeV Higgs boson confirms  the Higgs mechanism for electroweak symmetry breaking. Together with the non-observation of new particles to date, the question of naturalness -- what stabilizes the Higgs mass? -- becomes even more pressing. Inspired by the lightness of pions in low-energy QCD, a large class of models proposes new spontaneously broken global symmetries above the weak scale, in which the Higgs boson  arises as a pNGB. Prominent examples  include little Higgs models \cite{ArkaniHamed:2001nc,ArkaniHamed:2002qx,ArkaniHamed:2002qy} and the holographic Higgs models \cite{Contino:2003ve,Agashe:2004rs}. By now these models are  generically referred to as ``composite Higgs" models \cite{Bellazzini:2014yua,Panico:2015jxa}. Testing this class of models is among the top priorities in current and future experimental programs in high energy colliders. One generic feature of the model is the existence of fermionic top partners responsible for cancelling the top quadratic divergent contribution in the Higgs mass. If such a cancellation mechanism is confirmed in the future, it would be a striking confirmation of composite Higgs models \cite{Chen:2017dwb}. However, there is a distinct class of models in which the top partner is neutral under QCD, the so-called neutral naturalness models \cite{Craig:2015pha} which originated from the twin Higgs model \cite{Chacko:2005pe}. In this case it is challenging, experimentally, to discover the top partner. In this work we will focus on another generic feature of a composite Higgs: the nonlinear interactions due to the pNGB nature of the Higgs.

Effective Lagrangians for pNGB Higgs models rely on the seminal works of Coleman, Callan, Wess and Zumino (CCWZ) \cite{Coleman:1969sm, Callan:1969sn}, which requires specifying a broken group $G$ in the ultraviolet and an unbroken group $H$ in the infrared. The Nambu-Goldstone bosons (NGBs) are then  ``coordinates" parameterizing the coset space $G/H$. In CCWZ, each $G/H$ gives a seemingly different effective Lagrangian, valid  below the scale of new symmetry breaking, which is at $\Lambda=4\pi f\agt 10$ TeV. In addition to the pNGB Higgs boson, there are typically other composite resonances, at $m_\rho=g_\rho f \agt 1$ TeV, whose presence and interactions are model-dependent. Below $m_\rho$, one often  performs a ``matching" of the CCWZ Lagrangian to the Strongly Interacting Light Higgs (SILH) Lagrangian \cite{Giudice:2007fh}, so as to facilitate comparison with observables. There are numerous pNGB Higgs models, each based on a different coset $G/H$ \cite{Bellazzini:2014yua,Panico:2015jxa}, resulting in seemingly different effective Lagrangians and predictions. For simplicity, collider studies are often based on the $SO(5)/SO(4)$ minimal composite Higgs model.

Later it was realized that the CCWZ Lagrangian can be reformulated using only infrared data, by imposing the Adler's zero condition \cite{Adler:1964um}, without referring to a target coset $G/H$ \cite{Low:2014nga,Low:2014oga}. Thus in composite Higgs models with the custodial symmetry, where the Higgs transforms as a fundamental representation $\mathbf{4}$ of the $SO(4)$ subgroup of a potentially larger unbroken group $H$, the nonlinear Lagrangian below the cutoff scale $\Lambda$ and above the resonance scale $m_\rho$ is universal. In this Letter we extend the universal nonlinearity below the $m_\rho$ scale and argue that the universality is preserved in the gauge sector, which allows us to present new and universal predictions of a pNGB Higgs in its couplings to electroweak gauge bosons and in triple gauge boson couplings (TGC). 
Presently we summarize our arguments and findings; the details will be presented in a forthcoming publication \cite{Low:2018}.

\noindent
{\bf Universal Nonlinearity.}--One major development in modern S-matrix program concerns the reconstruction of effective field theories from the soft limits of scattering amplitudes. One prime example is the nonlinear sigma model (NLSM) describing interactions of NGBs, whose scattering amplitudes have long been know to possess the Adler's zero. Using a 4-point (pt) vertex as the building block, one can construct the 6-point tree-level amplitudes using only 4-pt vertices, which would not satisfy the Adler's zero condition. This necessitates the introduction of a 6-pt vertex whose value is uniquely determined by requiring the Adler's zero condition in the 6-pt amplitude. This program was initiated in Ref.~\cite{Susskind:1970gf} and completed to an arbitrary number of external legs in Refs.~\cite{Cheung:2014dqa,Cheung:2016drk}. At the level of the Lagrangian, the Adler's zero condition is equivalent to imposing constant shift symmetries in the NGB field, $\pi^a \to \pi^a +\vep^a+\cdots$, where $\vep^a$ is a set of infinitesimal constants. It turned out this is sufficient to reconstruct the NLSM Lagrangian without recourse to CCWZ and without reference to a coset \cite{Low:2014nga,Low:2014oga}. The equivalence of the nonlinear shift symmetry approach to that of the CCWZ becomes manifest once a suitable basis of group generators for $H$ is chosen.

Consider a set of scalars $\pi^a$ furnishing a linear representation of the unbroken group $H$, $\pi^a \to \pi^a + i \alpha^i (T^i)_{ab} \pi^b + {\cal O}(\alpha^2)$, where $T^i$ is the generator of $H$. We choose a basis where $T^i$ is purely imaginary and anti-symmetric, $(T^i)^T=-T^i$ and $(T^i)^{*}=-T^i$. It will be convenient to define the matrix ${\cal T}$:
\be
{\cal T}_{ab} =\frac2{f^2} (T^i)_{ar}(T^i)_{sb}\, \pi^r \pi^s \ ,
\ee
where $f$ is the NGB decay constant. Notice that we have changed the normalization of $f$ from that in Refs.~\cite{Low:2014nga,Low:2014oga,Low:2017mlh,Low:2018acv}, so as to conform with the convention in the literature in composite Higgs models. Then the nonlinear shift symmetry that enforces the Adler's zero condition to all orders in $1/f$ is \cite{Low:2017mlh,Low:2018acv}
\be
 \pi^{a\,\prime} = \pi^a + [F_1 (\mt)]_{ab}\  \vep^b\ ,\quad {F}_1({\cal T})= \sqrt{{\cal T}}\cot\sqrt{{\cal T}} \ , \label{eqshift}
\ee
and the leading order two-derivative Lagrangian invariant under the nonlinear shift symmetry is
\be
 {\cal L}^{(2)} = \frac12 [F_2(\mt)^2]_{ab} \ \partial_\mu\pi^a \partial^\mu\pi^b\ , \quad  {F}_2({\cal T})=\frac{\sin\sqrt{{\cal T}}}{\sqrt{{\cal T}}} \label{eqnlsmlagep}\ .
 \ee
Notice that Eq.~(\ref{eqnlsmlagep}) is written entirely using infrared data: the Adler's zero condition and the linearly realized group $H$. 
The broken group $G$ only serves the purpose of counting the number of NGBs, which is related to the normalization of the decay constant $f$. Applying Eq.~(\ref{eqnlsmlagep}) to the case of $\mathbf{4}$ under $SO(4)$ gives the universal NLSM Lagrangian for a composite Higgs.

So far we have only established the universal nonlinearity below the cutoff scale $\Lambda$ and above the resonance scale $m_\rho$. 
Integrating out the resonances results in the SILH Lagrangian. The crucial observation of SILH is that, after the composite resonances are integrated out,  {\em the nonlinearity of the effective Lagrangian is preserved}  \cite{Giudice:2007fh}. After the matching, one simply replaces the well-known $4\pi$ counting in NLSM by $g_\rho = m_\rho /f$. Operators in NLSM that are suppressed by $\Lambda = 4\pi f$ is now suppressed by $m_\rho=g_\rho f$ in the SILH Lagrangian. Otherwise the structure of the effective Lagrangian, including the nonlinearity, remains the same.  More explicitly, the NLSM Lagrangian is organized by the naive dimensional analysis (NDA) \cite{Manohar:1983md}, 
\be
S_{eff} = \int d^4 x \, \Lambda^2 f^2 \, {\cal L}\left(\frac{\pi}{f}, \frac{\partial}{\Lambda}\right) =  \int d^4x\,  {\cal L}^{(2)}+ \cdots  \ .
\ee
SILH replaces $\Lambda \to m_\rho$, but the nonlinear structure remains the same. Therefore  the ``symmetry-preserving" part of the SILH Lagrangian inherits the universal nonlinearity from the NLSM Lagrangian.

There are effects  that break the nonlinear shift symmetry of the NLSM. One example is the gauging of the electroweak $SU(2)_L\times U(1)_Y$ inside the $SO(4)$, which  can be incorporated by replacing the ordinary derivative, $\partial_\mu \to D_\mu = \partial_\mu + i  A_\mu$. Therefore gauging the electroweak symmetry doesn't spoil the universal nonlinearity in the NLSM. Generally speaking, massive spin-1 vectors coupling  to the Higgs current must be in the adjoint representation of $SO(4)$ \cite{Low:2009di}, and integrating out these massive vectors produce the same  nonlinear interactions as in the NLSM Lagrangian \cite{Giudice:2007fh}. This implies integrating out massive spin-1 vectors will also preserve the universality in SILH. It is possible to produce higher dimensional operators  violating the shift symmetry that are not present in the gauged NLSM Lagrangian. These operators, however, must be suppressed by the same coupling associated with the renormalizable interaction in the SM Lagrangian \cite{Giudice:2007fh}, such as the top Yukawa coupling and the quartic Higgs coupling. Therefore these operators reside in sectors containing fermions and Higgs potential energy. If we focus on the gauge sector, these violations of universal nonlinearity do not arise.


\noindent
{\bf ${\cal O}(p^2)$ Nonlinearity}.--After gauging the $SU(2)_L\times U(1)_Y$ subgroup of $SO(4)$, ${\cal L}^{(2)}$ in the unitary gauge becomes
\begin{align}
{\cal L}^{(2)} &= \frac{1}{2} \partial_\mu h \partial^\mu h+ \frac{g^2f^2}{4} \sin^2( \theta + h/f)   \nonumber \\
& \qquad \times \left(W^+_\mu W^{- \mu} + \frac{1}{2\cos^2 \theta_W} Z_\mu Z^\mu \right)\nonumber\\
&=\frac{1}{2} \partial_\mu h \partial^\mu h+ \left[1 + 2 \sqrt{ 1-\xi}\, \frac{h}{v} + (1-2\xi)\, \frac{h^2}{v^2} + \cdots\right]\nonumber\\
&\qquad\times \left( m_W^2  W^+_\mu W^{-\mu} + \frac{1}{2} m_Z^2 Z_\mu Z^{\mu} \right)\ ,
\end{align}
where $m_W = {g v}/{2}=m_Z {\cos \theta_W}$, $\sin\theta\equiv v/f$ and $\xi\equiv v^2/f^2=\sin^2\theta$. In particular, $v=246$ GeV is different from $\langle h\rangle =f \theta$, the vacuum expectation value of the pNGB Higgs $h$. It is important to note that, at this order, couplings of $n$ Higgs bosons with $WW$ and $ZZ$ are completely determined by 1) Adler's zero condition and 2) the Higgs boson sits in the $\mathbf{4}$ of the $SO(4)$. If we define the coefficient of $(h/v)^n (m_W^2 W_\mu^+ W^{-\, \mu}+m_Z^2 Z_\mu Z^\mu/2)$ to be $b_{nh}$, we see the first two coefficients are
\be
\label{eq:notest1}
b_{h}=2\sqrt{1-\xi} \ , \qquad b_{2h}=1-2\xi \ .
\ee
As have been emphasized, the normalization of $f$, and hence $\xi$, is not universal and dependent on the particular coset $G/H$. Thus we can extract $b_{h}$ from the $hVV$, $V=W,Z$, couplings and use it as an input to predict $b_{2h}$, which would be one experimental test of the universal nonlinearity. $b_{2h}$ could in principle be measured in $VV\to hh$ scattering process.

Since the coupling $b_{nh}$ is completely fixed by one free parameter $\xi$, there are many more predictions from the universality. It is worth stressing that, experimentally, the overarching goal should be to {\em overconstrain} the universality and perform as many tests as possible. However, at this order, all other predictions involve $b_{nh}$, $n\ge 3$, and require measuring $VV\to nh$, $n\ge 4$, which is a daunting task even in a future high energy collider. (The amplitude for $VV\to 3h$ vanishes for a symmetric coset  \cite{Contino:2013gna,Low:2014oga}.) It would be  more desirable if we could use only $2\to 2$ processes, which is possible if we include the ${\cal O}(p^4)$ operators.



\noindent
{\bf ${\cal O}(p^4)$ Nonlinearity}.--Turning off any gauging for now, there are two building blocks for ${\cal O}(p^4)$ operators: $d_\mu^a$ and $E_\mu^i$, which transforms as the $\mathbf{4}$ and $\mathbf{6}_A$ (adjoint) of $SO(4)$, respectively,
\bea
d_\mu^a(\pi, \partial) &=& \frac{\sqrt{2}}{f}[ {\cal F}_2({\cal T})]_{ab}\partial_\mu\pi^b \ , \nonumber \\
{E}_\mu^i(\pi, \partial) &=& \frac2{f^2}  \partial_\mu \pi^a [{\cal F}_4({\cal T})]_{ab} (T^i\pi)^b  \ ,
\eea
where ${\cal F}_2(\cal T)$ is defined in Eq.~(\ref{eqnlsmlagep}) and
\be
  {\cal F}_4({\cal T})=-\frac{2i}{\cal T}\sin^2\frac{\sqrt{\cal T}}2 \ .
\ee
In our notation $a, b, \cdots=1, \cdots, 4$ run in the $\mathbf{4}$, while $i,j,\cdots=1,\cdots, 6$ run in the $\mathbf{6}_A$, of $SO(4)$. Under the nonlinear shift symmetry in Eq.~(\ref{eqshift}), 
\bea
d_\mu &\to& U \ d_\mu \ , \nonumber\\
 E_\mu^i T^i &\to& U (E_\mu^i T^i) U^{-1} - i U \partial_\mu (U^{-1})\ ,
 \eea
where  $U\in H$ and its explicit form  can be found in Refs.~\cite{Low:2014nga,Low:2014oga}.

Gauging the electroweak $SU(2)_L\times U(1)_Y$ amounts to replacing $\partial_\mu \to D_\mu=\partial_\mu +i A_\mu $ in $d_\mu^a(\pi, \partial)$ and ${E}_\mu^i(\pi, \partial)$. Formally we can choose to gauge the full $SO(4)$ and $d_\mu^a$ now becomes
\be
d_\mu^a(\pi,D)=  \frac{\sqrt{2}}{f}  [ {\cal F}_2({\cal T})]_{ab} (D_\mu \pi)^b \ , \ D_\mu=\partial_\mu  + i A_\mu^i T^i \ ,
\ee
where $T^i$ is the generator of $SO(4)$. Although the gauging explicitly breaks the nonlinear shift symmetry, one can formally treat the gauge field as a ``spurion" in the $\mathbf{4}$ of $SO(4)$ that transforms covariantly in the same way as $d_\mu^a(\pi, \partial)$. This suggests a new building block that is covariant under both the shift symmetry and local gauge transformation,
\be
(f_{\mu\nu}^-)^a =  \frac{\sqrt{2}i}{f}  [ {\cal F}_2({\cal T})]_{ab} (T^i \pi)^b\, F_{\mu\nu}^i \ ,
\ee
where $F_{\mu\nu}^i$ is the field strength tensor of $A_\mu^i$. Similarly, from $E_\mu^i(\pi,D)$ one can identity another spurion in the $\mathbf{6}_A$ of $SO(4)$ and construct another covariant object:
\be
(f_{\mu\nu}^+)^i = F_{\mu \nu}^i + \frac2{f^2}F_{\mu\nu}^j (T^j \pi)^a [ {\cal F}_4({\cal T})]_{ab}(T^i\pi)^{b} \ ,
\ee
which comes from singling out the term in $E_\mu^i(\pi,D)$ proportional to $A_\mu^i$ and replacing the gauge field by its field strength tensor. Again both operators are constructed using only the infrared data, without recourse to a coset $G/H$. The corresponding operators in the CCWZ formalism can be found in Ref.~\cite{Contino:2011np}, whose notation we follow.

In the end there are four building blocks, $d^a_\mu, E_\mu^i, (f_{\mu\nu}^-)^a$ and $(f_{\mu\nu}^+)^i$, for constructing the ${\cal O}(p^4)$ operators. For objects carrying the $SO(4)$ adjoint index it is useful to further classify them as reducible representations of $SO(4)\sim SU(2)_L\times SU(2)_R$, an $SU(2)\times U(1)$ subgroup of which is identified with the electroweak gauge group. Their expressions in the unitary gauge, to all orders in $1/f$, are
\bea
 d_\mu^{{a}} &= &\sqrt{2} \left[{\delta^{{a} 4}}\, \partial_\mu\left( \frac{h}f\right)
+ \frac{ \delta^{ar}}{2} \sin(\theta+h/f)\right.\nonumber\\
&&\left.\phantom{\frac12}\quad\times  (W_\mu^r - \delta^{r3} B_\mu)\right]\ , \nonumber \\
 (E_{\mu}^{L/R})^r &=& \frac{1 \pm \cos (\theta+h/f) }{2} W_\mu^r \nonumber\\
 && \quad+ \frac{1\mp \cos(\theta+h/f)}{2} B_\mu \delta^{r3}\ ,\nonumber \\
(f_{\mu \nu}^-)^{{a}}&=&\frac1{\sqrt{2}}  \sin(\theta+h/f)  (W_{\mu \nu}^r - \delta^{r3} B_{\mu \nu}) \delta^{ra},\nonumber \\
(f_{\mu \nu}^{+L/R})^{r} &=& \frac{1\pm \cos (\theta+h/f)}{2} W^r_{\mu \nu} \nonumber \\
&&\quad + \frac{1\mp \cos (\theta+h/f)}{2} \delta^{r3} B_{\mu \nu} 
\eea
where the superscripts $L$ and $R$ refer to the upper and lower signs, respectively, and $r=1,2,3$ is the adjoint index in $SU(2)_{L/R}$. 
 There are 11 independent operators at ${\cal O}(p^4)$ that can be constructed \cite{Contino:2011np}, six of which are even under space inversion $\vec{x}\to -\vec{x}$ and not contracted with $\epsilon_{\mu\nu\rho\sigma}$. We focus on these six CP-even operators in this work and compute them to all orders in $1/f$.  They are:
\begin{align}
\label{eq:op4}
O_1 & =  \left( d_\mu^{{a}}d^{\mu a} \right)^2,\  \quad O_2 =  (d_\mu^{{a}} d_\nu^{{a}})^2\ , \nonumber \\
O_3 &= \left[\left(E_{\mu \nu}^{L} \right)^r \right]^2- \left[\left(E_{\mu \nu}^{R} \right)^r \right]^2\ ,\nonumber  \\
O_4^\pm &= -i\, d_\mu^a d_\nu^b \left[ (f^{+L}_{\mu \nu})^r\, T_L^{r} \pm (f^{+R}_{\mu \nu})^r\, T_R^{r} \right]_{ab}\ ,\nonumber  \\
O_5^+&=\left[(f_{\mu \nu}^-)^{{a}}\right]^2  ,\  \ \  O_5^-=\left[(f_{\mu \nu}^{+L})^{r}\right]^2-\left[(f_{\mu \nu}^{+R})^{r}\right]^2 ,
\end{align}
where $T_{L/R}^r$ is the $SU(2)_{L/R}$ generator. Using the SILH power counting,  the four-derivative effective action is:
\be
\label{eq:lag4}
S_{SILH}^{(4)}= \int d^4x \ m_\rho^2\, f^2 {\cal L}^{(4)}\left(\frac{\pi}f, \frac{D}{m_\rho}\right)=\sum_i \frac{c_i}{g_\rho^2} O_i \ ,
\ee 
where $c_i$ are expected to be order unity constants, although in some cases operators  contributing to  couplings of neutral particles and the on-shell photon  are further suppressed by additional loop factors \cite{Giudice:2007fh}.


\begin{table}[!t]
\footnotesize
\begin{center}
\begin{tabular}{|l|c |c| c| c| c|c|c|}
\hline
  $\mI^{h}_i$ &   $C^h_i$\\
 \hline
(1) $  {h} Z_{\mu} \ewd^{\mu \nu} Z_{\nu}/v$ & $ [{4 c_{ 2 w} }  (-2 c_3  + c_4^- )  
+{4}  c_4^+\cos \theta]/c^2_{w}$ \\
 \hline
(2)  $ {h}  Z_{\mu\nu} Z^{\mu\nu}/v$ & $ -2[   ( c_4^+  - 2 c_5^+ ) \cos \theta +  ( c_4^-  + 2 c_5^- ) c_{2w}  ]/ {c^2_{w}} $\\
 \hline
(3)  $ {h}  Z_{\mu} \ewd^{\mu \nu} A_{\nu}/v$ & $8  ( - 2 c_3  +   c_4^- )  t_{w} $ \\
\hline
(4) $ {h} Z_{\mu\nu} A^{\mu\nu}/v$ & -$4  (  c_4^-   +2 c_5^- )  t_{w} $\\
\hline
\hline
  $\mI^{2h}_i$ &  $C^{2h}_i$  \\
 \hline
(1) $  {h^2} Z_{\mu} \ewd^{\mu \nu} Z_{\nu}/{v^2}$ & $ 2 [{c_{ 2w} }( - 2 c_3  +  c_4^-  ) \cos \theta    +    c_4^+ \cos 2 \theta ]/{c^2_{w}} $  \\
 \hline
(2) $  {h^2} Z_{\mu\nu} Z^{\mu\nu}/{v^2}$ & $ -[   (c_4^+ - 2 c_5^+ ) \cos 2 \theta  + {c_{2 w}}  (c_4^-  + 2 c_5^-  ) \cos \theta  ]/{c^2_{w}}  $ \\
 \hline
(3) $  {h^2} Z_{\mu} \ewd^{\mu \nu} A_{\nu}/{v^2}$ & $ 4  t_{w}  ( -2c_3   + c_4^- ) \cos \theta $  \\
 \hline
(4) $  {h^2} Z_{\mu\nu} A^{\mu\nu}/{v^2}$ & $ -2  t_{ w}   (   c_4^-  + 2 c_5^-  )  \cos \theta $  \\
\hline
\hline
  $\mI^{3V}_i$ &   $C_i^{3V}$  \\
 \hline
$  i\, g\, c_{w} W^{+\mu \nu}  W^-_\mu  Z_\nu $ & $ \delta \tilde{g}_1^Z=  2[   ( 2c_3 - c_4^- ) \cos \theta -    c_4^+ ]/{c^2_{w}}   $  \\
\hline
$i\, g\,  c_{w}W^{+}_{\mu} W^{-}_{ \nu}   Z^{\mu \nu} $& $   \delta\tilde{ \kappa}_Z  ={2}[ (2 c_3 - c_4^- ) \cos \theta  -   c_4^+ c_{2w}   - 4 c_5^+ s^2_{w} ]/{c^2_{w}}   $   \\
\hline
$i\, e\, W^{+}_{\mu} W^{-}_{ \nu} A^{\mu \nu}  $& $ \delta\tilde{ \kappa}_\gamma =- 4(  c_4^+  - 2  c_5^+ )    $  \\
\hline
\end{tabular}
\end{center}
\caption{Predictions of universal Higgs nonlinearity. Here $c_w=\cos\theta_W$ and $c_{2w}=\cos2\theta_W$, where $\theta_W$ is the weak mixing angle, and  $t_w=\tan\theta_W$ .}
\label{tab:coupling}
\end{table}

\noindent
{\bf Predictions}.--Here we present predictions of nonlinear interactions in Higgs couplings to electroweak gauge bosons and TGC in the unitary gauge.  We start by parameterizing the couplings as follows: 
\beq
\label{eq:Cis}
\mL_h =  \sum_i \frac{m_W^2}{m_\rho^2} \left(C^h_i \mI^h_i  +C^{2h}_i \mI^{2h}_i +  C^{3V}_i \mI^{3V}_i  \right) + {\rm h.c.} \ ,
\eeq
where $\mI_i^{h}$ and $\mI_i^{2h}$ are ${\cal O}(p^4)$ operators contributing to $hVV$ and $hhVV$ couplings, respectively, while $\mI^{3V}_i$  involves TGC.  Here $V=W,Z,\gamma$. The $C_i$'s in the above are dimensionless coefficients. We list only a subset of the operators  and predictions from the universal nonlinearity in Table \ref{tab:coupling}. The complete list and prediction will be presented in Ref.~\cite{Low:2018}. Notice that $O_{1,2}$ in \Eq{eq:op4} do not contribute to operators appearing in  Table \ref{tab:coupling}. 


Recall that  $c_i$'s in Eq.~(\ref{eq:lag4}) are uncalculable coefficient in the derivative expansion of the effective theory and parameterize our ignorance of the ultraviolet physics. From Table \ref{tab:coupling} we can extract  "universal relations" among $C_i$'s in Eq.~(\ref{eq:Cis}) that are independent of the  $c_i$'s. Some examples are
\bea
\label{eq:ident1}
&&\frac{C^{2h}_3}{C^h_3} = \frac{C^{2h}_4}{C^h_4} = \frac{ 2c^2_{w}  C^{h}_2 -c_{2w} C^{h}_4/t_{w} }{2\delta\tilde{\kappa}_\gamma}\nonumber \\
&&\qquad\qquad=\frac12 \cos\theta =\frac12\left(1-\frac12 \xi+\cdots \right) \ ,\\
&&\frac{2c^2_w C^{2h}_1 - c_{2w}   C^{2h}_3/ t_w} {2c^2_w C^h_1-c_{2w}   C^{h}_3/ t_w} =\frac{2c^2_w C^{2h}_2 - c_{2w}   C^{2h}_4/ t_w} {2c^2_w C^h_2-c_{2w}   C^{h}_4/ t_w} \nonumber\\
&& \qquad\qquad = \frac{\cos2\theta}{2\cos\theta}= \frac12\left(1 - \frac32 \xi + \cdots\right)\ ,
\label{eq:ident2}\\
&&\delta \tilde{\kappa}_Z = \delta \tilde{g}_1^Z - t_w^2\, \delta \tilde{\kappa}_\gamma   \ ,
\label{eq:ident3}
\eea
where the notation is explained in the caption of Table  \ref{tab:coupling}. These relations are all determined by the one single input parameter $\sin\theta$ and free from the uncalculable coefficients in Eq.~(\ref{eq:lag4}). As such they serve as universal predictions of Higgs nonlinearity and can be tested in future experimental programs on the Higgs boson. Note that $\mI_3^{h,2h}$ do not contribute to processes involving the on-shell photon. 


It is interesting to compare the prediction of universal nonlinearity with theories without a pNGB Higgs, e.g. the Standard Model Effective Field Theory (SMEFT) \cite{Henning:2014wua} which augment the SM Lagrangian with higher dimensional operators with arbitrary coefficients. In particular, the dim-6 and dim-8 operators relevant for our discussion are parameterized as follows:
\be
{\cal L}_{SMEFT} \supset \sum_{i=W,B,HW,HB} \frac{\alpha_{i}}{m_\rho^2}\,  \mO_{i} +   \frac{\alpha_{i}^8}{f^2 m_\rho^2} \, (H^\dagger H)\mO_{i}\ ,
\ee
where $\mO_W,\mO_B,\mO_{HW},\mO_{HB}$ are defined explicitly in Ref.~\cite{Giudice:2007fh} with the gauge boson fields canonically normalized. 
Then Eqs.~(\ref{eq:ident1}) and (\ref{eq:ident3}) now become, at the leading order in $\xi$,
\bea
&&\frac{C^{2h}_3}{C^h_3}\approx  \frac12 \left(1 + \frac{\alpha_{HW}^8 - \alpha^8_{HB}}{\alpha_{HW} - \alpha_{HB}}\xi \right),\\
&& \frac{C^{2h}_4}{C^h_4} \approx \frac12 \left(1 + \frac{\alpha_{W}^8 - \alpha^8_{B} + \alpha_{HW}^8 - \alpha^8_{HB} }{\alpha_W - \alpha_B + \alpha_{HW} - \alpha_{HB}}\xi\right),\\
&&\frac{2c^2_w C^{2h}_1 - c_{2w}   C^{2h}_3/ t_w} {2c^2_w C^h_1-c_{2w}   C^{h}_3/ t_w} \nonumber\\
&&\qquad \approx\frac12 \left(1 + \frac{\alpha_{W}^8 + \alpha^8_{B} + \alpha_{HW}^8 + \alpha^8_{HB} }{\alpha_W +\alpha_B + \alpha_{HW} +\alpha_{HB}}\xi\right),\\
&&\frac{2c^2_w C^{2h}_2 - c_{2w}   C^{2h}_4/ t_w} {2c^2_w C^h_2-c_{2w}   C^{h}_4/ t_w} \nonumber\\
&&\qquad \approx \frac12 \left(1 + \frac{ \alpha_{HW}^8 + \alpha^8_{HB} }{\alpha_{HW} + \alpha_{HB}}\xi\right).
\eea
Comparing these relations, we can see that the difference between the universal nonlinearity and SMEFT start from $\mO(\xi)$, which is the contribution from the dim-8 operators. This is to be expected, since at the dim-6 level the nonlinearity makes no prediction due to the arbitrary normalization of $f$. However, the ratio of the dim-8 coefficient with that of the dim-6 is determined by the nonlinearity. The SMEFT, on the other hand, makes no assertion about the coefficients at any order. However, it is interesting to observe that \Eq{eq:ident3} holds in SMEFT at the dim-6 level and that it is preserved by the Higgs nonlinearity to all orders in $1/f$. 
\noindent
{\bf Conclusion}.-- In this work we argued for the universal nonlinearity of models containing a pNGB Higgs boson that transforms as the $\mathbf{4}$ of an $SO(4)$ subgroup of the unbroken group $H$, without referring to a particular coset space $G/H$. 
We presented the effective action up to ${\cal O}(p^4)$, as well as its predictions in the Higgs couplings to electroweak gauge bosons and the TGC. These universal relations are fixed by a single input parameter $\sin\theta$, which is related to the normalization of $f$, and could potentially be tested using multivariate techniques in collider experiments \cite{Chen:2014pia,Gainer:2014hha,Craig:2015wwr,Brehmer:2016nyr}. In addition, the TGC could also be probed to a high precision  at the HL-LHC~\cite{Franceschini:2017xkh,Liu:2018pkg}. In a forthcoming publication \cite{Low:2018} we will provide a comprehensive phenomenological analysis of the universal relations, which would serve as the smoking gun signal of the pNGB nature of the Higgs boson. 

 This work is supported in part by the U.S. Department of Energy under contracts No. DE-AC02-06CH11357 and No. DE-SC0010143.  



\bibliography{references_Higgs}



\end{document}